\documentclass{raa}            % referee version: for submission

\usepackage{graphicx,times}             %for PS/EPS graphics inclusion, new
\usepackage{natbib}
\usepackage{amssymb,amsmath}

\usepackage[authormarkup=superscript, final]{changes} % 加载宏包
\definechangesauthor[name={Your Name}, color=blue]{me}
\setdeletedmarkup{\textcolor{red}{\sout{#1}}}
\setaddedmarkup{\textcolor{blue}{#1}}

\usepackage{array}
\usepackage{booktabs}
\usepackage{caption}
\usepackage{float}

\usepackage{siunitx} % 专业单位排版

\DeclareSIUnit{\pixel}{pix}
\DeclareSIUnit\arcminute{arcmin}

\bibpunct{(}{)}{;}{a}{}{,}

\usepackage[pagebackref=true]{hyperref}

\begin{document}

  \title{Impact of Satellite Constellations on Observations with the 80-cm Telescope and the Mini-SiTian at the Xinglong Observatory, NAOC 
}
%   \subtitle{I. Place Your Subtitle Here}

   \volnopage{Vol.0 (20xx) No.0, 000--000}      %%preserved for Editor. DOn't remove!
   \setcounter{page}{1}          %%starting page, preserved for Editor. DOn't remove!

   \author{Jing Ren %(周爱英) %% Put your Chinese name in "( )" if you like. Note to open line 11 "\usepackage[UTF8]{ctex}"
      \inst{1,2}
   \and Zhou Fan
      \inst{1,2,6}
   \and Hong-rui Gu
      \inst{1,2}
   \and Qi-qian Zhang
      \inst{1,2}
   \and Yun-fei Xu
      \inst{1,2,3}
    \and Jun-ju Du      
      \inst{4}
   \and Xiao-han Chen
      \inst{1,5}
   \and Lin-ying Mi
      \inst{1,2,3}
   \and Hong Wu
      \inst{1,2}
   }

   \institute{National Astronomical Observatories, Chinese Academy of Sciences, Beijing 100101, People’s Republic of China; {\it zfan@nao.cas.cn}\\
        \and
             School of Astronomy and Space Science, University of Chinese Academy of Sciences, Beijing 100049, People’s Republic of China\\
        \and
             National Astronomical Data Center of China, Beijing 100101, People’s Republic of China\\
        \and
             Shandong Key Laboratory of Space Environment and Exploration Technology, Institute of Space Sciences, School of Space Science and Technology, Shandong University, Shandong, China\\
        \and             
             School of Physics and Astronomy, China West Normal University, Nanchong 637002, People’s Republic of China\\
        \and
             Corresponding author: Zhou Fan(zfan@nao.cas.cn)\\
\vs\no
   {\small Received 20xx month day; accepted 20xx month day}}

\abstract{The \added{rapid} development of \deleted{LEO} mega-constellations \added{in low Earth orbit (LEO)} severely impacts ground-based optical astronomical observations. \added{By combining WorldWide Telescope (WWT) simulations with 2019 and 2023 observational data from the Xinglong Observatory 80-cm telescope and 2023 data from the Mini-SiTian (MST), we find that satellite visibility increases with deployment, particularly during the summer. For the 80-cm telescope, the fraction of images containing satellite trails increased from an average of 0.34\% in 2019 to 0.7\% in 2023; meanwhile, for the MST in 2023, the fraction rose from 5\% in January to 12\% by December, peaking at 19\% in the summer. Through stratified analysis of solar elevation and local time, we find that observations during twilight and summer are particularly susceptible to satellite trail interference.} \added{Photometric analysis reveals that the interference intensity increases for fainter sources and those closer to the trails. Furthermore, a comparative analysis across different seeing conditions shows that the deviation of median standardized residuals ($\sigma$) is significantly greater under poor seeing than under good seeing conditions.}
}

   \authorrunning{Jing Ren et al.}            %author_head in even pages
   \titlerunning{Impact of Satellite Constellations on observations at the Xinglong Observatory }  % title_head in odd pages

   \maketitle
%% The author head (on even pages) and the title head (on odd pages) will be
%% automatically extracted from \author{} and \title{}. Whenever the title is too long,
%% you will be asked to supply a shorter one by inserting either \authorrunning{} or
%% \titlerunning{} before \maketitle. Anyway, you can specify your own heads.
%%
%%
%% Note: In the following text body of your manuscript, please note several differences from
%%       other major journals:
%% (1) \subsection{Please Capitalize the First Letter of Each Notional Word in Subsection Title}
%% (2) Please Capitalize the First Letter of Each Notional Word in all tables' captions

%
%________________________________________________ sections below
%
\section{Introduction}           %% first-level sections will be auto-capitalized
\label{sect:intro}
Since the launch of the first artificial satellite in 1957, the region of space near Earth has undergone a radical transformation. A key driver of \added{this} recent change is the proliferation of satellite mega-constellations—vast networks of satellites. Their deployment has led to a rapid expansion \added{of the satellite population} in LEO, with the number of active satellites more than doubling since early 2019 \citep{barentine2023aggregate}. \deleted{Today,} \added{Currently,} mega-constellation satellites constitute an emerging \added{source of} \deleted{contributor to } astronomical light pollution. In recent years, \deleted{the} rapid \deleted{advancement} \added{advancements} in space communication \deleted{technology} \added{technologies} and \deleted{the} significant \deleted{reduction} \added{reductions} in satellite launch costs have sparked a \deleted{new boom} \added{rapid acceleration} in the development of LEO satellite constellations. This development is trending toward \deleted{large-scale deployment} \added{massive, large-scale deployments} \citep{muirhead2025modeling}. The Starlink project, leveraging the rapid growth of SpaceX's space industry, exemplifies this trend. SpaceX has received approval from the U.S. Federal Communications Commission (FCC) to operate up to 12,000 satellites. Subsequently, the company filed an application with the FCC for an additional 30,000 Starlink satellites, bringing the total planned constellation size to 42,000 \citep{halferty2022photometric}. Since the launch of the first batch of 60 Starlink satellites by SpaceX in May 2019, the astronomical community has persistently expressed concerns about the potential adverse impacts these satellites may have on ground-based astronomical observations \citep{mcdowell2020low}. Other companies, such as OneWeb, Amazon, and Samsung, along with national agencies, are also planning similar LEO satellite constellations. Currently, there are 43 mega-constellation programs still in the development phase\citep{zhi2024multicolour}. According to regulatory filings with the International Telecommunication Union \added{(ITU)}, approximately 100,000 satellites are projected to be launched into LEO within the next decade \citep{krantz2021characterizing}. Table~\ref{tab:constellations} lists some information for frequently mentioned mega-constellations. \deleted{of artificial satellites.}

\citet{gallozzi2020concerns} \deleted{quantifying} \added{quantified} ground-based \deleted{astronomy damages, due to} \added{astronomical damage; they argued that due to} satellite disruption, the loss in \added{the} value of \deleted{the} public investment in each \deleted{ground-based astronomical} facility varies directly with the loss of \added{the} associated scientific \deleted{content from its observations} \added{output}. Data contaminated by satellite interference \deleted{often must be discarded} \added{frequently require exclusion}, thereby eroding the return on investment for \deleted{expensive research facilities and the efforts to mitigate such issues} \added{both high-cost research facilities and the subsequent mitigation efforts} \citep{kruk2023HST}. These losses amount to hundreds of millions in public funds annually, \deleted{accumulating to} \added{potentially totaling} tens of billions over \deleted{decades} \added{the coming decades}. If the number of satellites in LEO reaches 100,000, this scenario would inflict incalculable damage not only \added{to} ground-based astronomical observations but also \added{to} space exploration as a whole \citep{gallozzi2020concerns}.

Multiple studies\added{---}including the Satellite Constellations 1 and 2 (SATCON1 and SATCON2) workshops and \added{the} Dark and Quiet Skies for Science and Society \added{workshops---}have investigated the impact of satellite constellations on astronomical observations. The SATCON1 report also incorporates \deleted{substantial recent observational data on} \added{extensive observations of} Starlink satellites. Collectively, these \deleted{observations} \added{data} confirm that \deleted{astronomers are justified in their concerns regarding} \added{concerns regarding the preservation of} the night sky \added{are well-founded}: \deleted{the} satellites currently in orbit \deleted{are bright} \added{exhibit high apparent brightness}. When illuminated by sunlight, many \added{units} are visible to the unaided eye, \added{particularly since} their brightness levels \deleted{are currently unregulated} \added{remain currently subject to no international regulatory limits} \citep{ref2}. Several astrophysical research domains are identified by \citet{walker2020impact} as critically vulnerable to the effects of large LEO satellite constellations, including observations of rare transients, deep \added{and} wide extragalactic imaging, searches for near-Earth objects (NEOs), \added{and} deep wide-field near-infrared (NIR) \deleted{imaging, among others} \added{surveys}. 

\citet{mcdowell2020low} surveyed the current population of artificial satellites in LEO and modeled the impacts of various \added{satellite} constellations on ground-based observatories at different geographic locations. \added{The study} \deleted{noting} \added{noted} particularly \deleted{pronounced effects for} \added{severe impacts on} twilight observations, long-exposure \deleted{exposures with wide fields of view} \added{wide-field exposures}, high-latitude observatories, and \added{observations during} \added{local} summer. \deleted{months.} \citet{ref2} analyzed \deleted{the} Zwicky Transient Facility (ZTF) data from November 2019 to September 2021 and reported a strong correlation between \added{the} increasing \deleted{SpaceX satellite deployments} \added{number of SpaceX satellites} and the frequency of image contamination. \deleted{twilight observations were most affected, with compromised frames increasing from less than 0.5\% to 18\%, leading to the projection that nearly all ZTF twilight exposures would contain trails once the Starlink constellation reaches 10,000 satellites} \added{Specifically, twilight observations were most severely impacted, with the fraction of compromised frames increasing from less than 0.5\% to 18\%. This trend led to the projection that nearly all ZTF twilight exposures will contain satellite trails once the Starlink constellation reaches 10,000 units.} \citet{bassa2022analytical} \deleted{simulated and analyzed} \added{performed comprehensive simulations to evaluate} the impact of satellite constellations on various astronomical instruments, \deleted{such as} \added{including both} imagers and spectrographs. \citet{lawler2021visibility} developed a satellite reflectance model calibrated with Starlink observations to \deleted{estimate visible magnitude and on-sky distribution across locations, seasons, and times} \added{predict apparent magnitudes and spatial distributions of satellites as a function of geographic coordinates, season, and time}. \added{The} results indicate that Starlink satellites at \added{an altitude of} 550 km remain persistently visible throughout summer nights at \deleted{mid-high} \added{mid-to-high} latitudes. \citet{tyson2020mitigation} analyzed the effects of \added{satellite constellations} on the \deleted{upcoming} Vera C. Rubin Observatory’s Legacy Survey of Space and Time (LSST), empirically determining that satellites \deleted{exceeding} \added{brighter than} $V \approx 7$ mag induce \added{nonlinear} CCD cross-talk artifacts.

In response to strong advocacy from the astronomical community, SpaceX has developed new designs aimed at reducing satellite reflectivity. These mitigation \deleted{approaches} \added{strategies} include \added{the use of} experimental coatings to diminish \added{the} albedo and fitting \deleted{them} \added{the spacecraft} with \deleted{sun visors} \added{deployable sunshades}. In 2020, SpaceX implemented \deleted{an experimental low-reflectivity coating on the DarkSat test satellite} \added{its initial mitigation attempt with "DarkSat" (Starlink-1130), which featured an experimental black darkening coating} \citep{tregloan2020first}. \citet{tregloan2020first} and \citet{halferty2022photometric} \deleted{found it to be dimmer} \added{reported a reduction in optical brightness,} but \citet{horiuchi2020simultaneous} reported that it was brighter. In either case, significant \deleted{sunlight absorption by the spacecraft} \added{solar absorption} caused thermal management issues, leading to the abandonment of this strategy. The \deleted{second approach} \added{subsequent iteration} involved the deployment of \added{the} "VisorSat" \deleted{Starlink satellites} \added{design}, which \deleted{are} \added{was} equipped with \deleted{sun visors} \added{deployable sunshades}. This modification \deleted{has} significantly reduced satellite brightness \citep{mallama2023assessment}. 

In this study, we assess the impact of artificial satellites on observations with the 80-cm telescope and the MST at the Xinglong Observatory by integrating simulation and observational approaches. Using the 
\deleted {WorldWide Telescope to simulate} \added{WWT, we simulate} Starlink satellite visibility \deleted{simulate} \added{and quantify} the number of \deleted{Starlink satellites that are} illuminated \added{spacecraft} \deleted{by the Sun} as a function of date and time of night. \deleted{Analyze} \added{We then analyze} observational data from \deleted{the 80-cm telescope and the MST} \added{both instruments}, employing the probabilistic Hough transform to detect satellite trails and \deleted{quantifying} \added{calculating} the fraction of images containing satellite trails.
Subsequently, we \deleted{quantified} \added{quantify} the fraction of sources affected by satellite trails. Furthermore, for \deleted{the} trail-contaminated images \deleted{captured} \added{acquired} by the MST, we \deleted{performed} \added{perform} aperture photometry to derive \deleted{the} standardized residuals between \deleted{our} measured magnitudes and \deleted{the} Gaia catalog magnitudes. The sources are then grouped by both their distance from the \added{satellite} trail and their magnitude to analyze the degree of impact on sources at different distances and with varying \deleted{brightness levels } \added{magnitudes}.

The paper is structured as follows: in Section \ref{sect:Simulation}, we \deleted{simulated} \added{simulate} the visibility of Starlink satellites using the \deleted{Worldwide Telescope} \added{WWT} to evaluate their impact on observations at the Xinglong Observatory. Section \ref{sect:trails} \deleted{applies} \added{employs} the probabilistic Hough transform to detect satellite trails \deleted{for the} \added{within the} 80-cm telescope \added{datasets} from 2019 and 2023, \deleted{and for the} \added{as well as} MST \added{data} from 2023, quantifying the fraction of images containing satellite trails. Section \ref{sect:photometry} \deleted{conducts photometry on} \added{presents a photometric analysis of} trail-contaminated MST images to assess the impact on sources at various distances from the trail and \deleted{with} \added{across} different brightness levels. Section \ref{sect:discussion} \deleted{is a discussion of this work and} \added{discusses these findings in the context of} current research. \added{Finally,} Section \ref{sect:conclusion} \deleted{is the summary and outlook of this work} \added{provides a summary and future outlook}.

\begin{table}[H]
\centering
\caption{Information of Several LEO Mega-constellations}
\label{tab:constellations}
\begin{tabular}{c>{\centering\arraybackslash}p{0.22\textwidth}
                >{\centering\arraybackslash}p{0.22\textwidth}
                >{\centering\arraybackslash}p{0.18\textwidth}
                >{\centering\arraybackslash}p{0.15\textwidth}}
\toprule
\textbf{No.} & \textbf{Constellation} & \textbf{Manufacturer} & \textbf{Altitude (km)} & \textbf{Satellites} \\
\midrule
1 & Starlink Generation 1 & SpaceX & 335-560 & 11,926 \\
2 & Starlink Generation 2 & SpaceX & 328-614 & 30,000 \\
3 & Project Kuiper & Amazon & 590-630 & 3,236 \\
4 & OneWeb Phase 1 & OneWeb/Airbus & 1200 & 1,980 \\
5 & OneWeb Phase 2 & OneWeb/Airbus & 1200 & 6,372 \\
6 & GuoWang GW-A59 & CASC & 508-1,145 & 12,992 \\
7 & Hanwha System & Hanwha & Unknown & 2,000 \\
\bottomrule
\end{tabular}
\end{table}

%% Authors can give a citation as 'Michel et al. 1992'.(\citealt{Michel+etal+1990, Michel+etal+1992})
%% You may also use \cite, \citep and \citet for citation, and use Table~1 or Figure~1
%% and so forth. Using \ref and \label for cross-references of Tables/Figures
%% is a good way in adjusting/adding/removing text, tables or figures.

\section{Simulation with the WorldWide Telescope}
\label{sect:Simulation}
We \deleted{simulated} \added{simulate} the visibility of Starlink satellites using the \deleted{WorldWide Telescope (}WWT\deleted{)} to assess their impact on observations at the Xinglong Observatory. \added{We} \deleted{simulated} \added{quantify} the number of Starlink satellites visible from \deleted{the Xinglong Observatory} \added{this site} \deleted{from January to December 2023, and separately for summer and winter, simulated the number of visible Starlink satellites} \added{throughout 2023, including targeted seasonal analyses for summer and winter} as a function of date and time.

\subsection{WorldWide Telescope}
The \deleted{WWT, an open-source platform for the visualization of astronomical data, was originally developed by Microsoft Research and is currently maintained by the American Astronomical Society} \added{WWT is an open-source astronomical data visualization platform, originally developed by Microsoft Research and now maintained by the American Astronomical Society} \citep{xu2020ivoa,rosenfield2018aas}. The WWT boasts powerful spatiotemporal simulation capabilities, enabling the simulation of celestial events for any specified time and location. Its Sats Tracker module simulates satellite visibility, while the Obs Simulator module loads and visualizes telescope observation plan files, simultaneously simulating the observed sky regions and satellite motion \citep{cui2022impact}.

\subsection{Impact of Starlink Satellites on observations at the Xinglong Observatory}
Based on the capabilities of the WWT, \deleted{using Starlink satellites as a representative case} \added{taking Starlink as a representative constellation}, we \deleted{simulated} \added{simulate} the impact of Starlink on the observable sky regions of the Xinglong Observatory \deleted{over a one-year period from January to December 2023} \added{throughout 2023}. The simulation \deleted{estimated} \added{quantifies} the number of Starlink satellites visible from the observatory. Furthermore, we \deleted{simulated} \added{simulate} the number of visible Starlink satellites as a function of date and time of night for both summer and winter. The Xinglong Observatory of \added{the} National Astronomical Observatories, CAS (NAOC), was founded in 1968, and it is \deleted{L} \added{l}ocated \deleted{at} \added{in} the south of the main peak of the Yanshan Mountains, in the Xinglong County, Hebei Province, which is $\sim\!120\,\mathrm{km}$ northeast of Beijing. Its geographical coordinates are 117$^\circ$34.5$^\prime$ E longitude and 40$^\circ$23$^\prime$36$^{\prime\prime}$ N latitude. The average altitude of the Xinglong Observatory is $\sim\!960\,\mathrm{m}$. It is one of the primary observing stations of NAOC. As one of the largest optical astronomical observatory sites in the continent of Asia, it harbors nine telescopes with an effective aperture greater than 50 cm \citep{fan2016}. 

We \deleted{Employing} \added{employ} the 1st and 15th of each month as designated sampling dates. For each date, we \deleted{computed} \added{compute} time windows \deleted{when} \added{during which} the solar elevation angle \deleted{was} \added{is} below $-12^\circ$, spanning from the end of evening nautical twilight to the \deleted{end} \added{beginning} of morning nautical twilight. This is because satellite visibility depends on the sunlight it reflects and the sky’s brightness. Only after the Sun is below a certain elevation angle can these two factors lead to a bright satellite. Therefore, we choose a Sun elevation angle of $-12^\circ$ as the limit \citep{cui2022impact}. \added{We retrieve} \deleted{The} Starlink satellite orbital data \deleted{was then retrieved} from the Space-Track website. This website publishes Starlink \deleted{satellite} orbital data in \added{Two-Line Element (}TLE\added{)} format, which is derived from \deleted{the} ephemerides provided by SpaceX and is accessible via a dedicated API. Subsequently, the Sats Tracker module \deleted{divided} \added{divides} the simulation period into 5-second time intervals, sequentially computing which satellites \deleted{would} appear within the observable sky region of the observatory during each interval. Following the computation of all time slices, the start and end times of each satellite's visible period \deleted{were} \added{are} determined. Furthermore, since most telescopes have a pointing limit of at least $30^\circ$ above the horizon, \added{detected} satellites with a maximum elevation below $30^\circ$ \deleted{were also discarded} \added{are excluded from the analysis}.

\begin{figure}[H]    
\centering    
\includegraphics[width=1\linewidth]{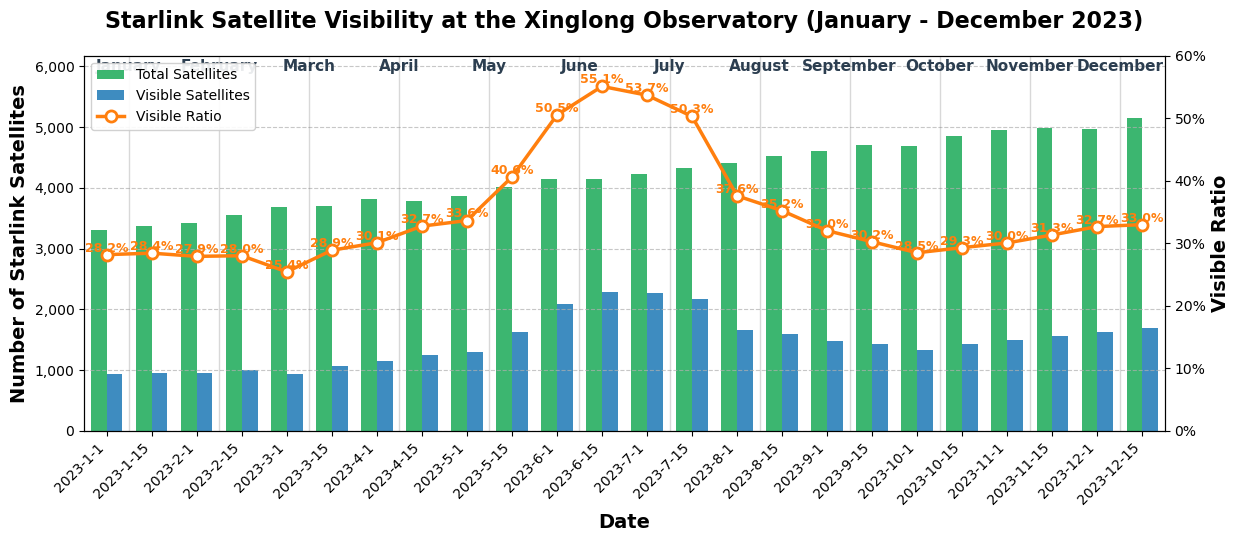}    
\caption{From January to  December 2023, the number of Starlink satellites in orbit and the number of Starlink satellites visible from the Xinglong Observatory. Green bars represent the total number of in-orbit Starlink satellites, blue bars indicate the number of visible Starlink satellites, and the orange line plots the ratio of visible to total satellites.}    
\label{fig:1}
\end{figure}

\begin{figure}[H]
\centering
\begin{minipage}{0.47\textwidth}
\centering
\includegraphics[width=\linewidth]{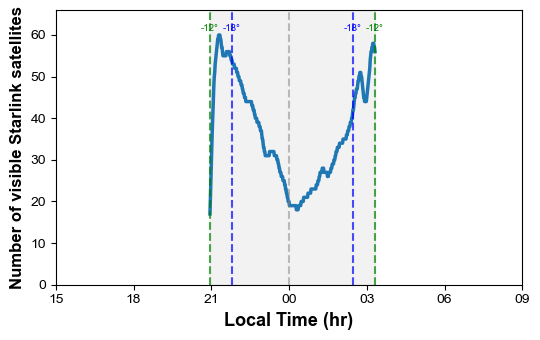}
\end{minipage}
\hfill
\begin{minipage}{0.47\textwidth}
\centering
\includegraphics[width=\linewidth]{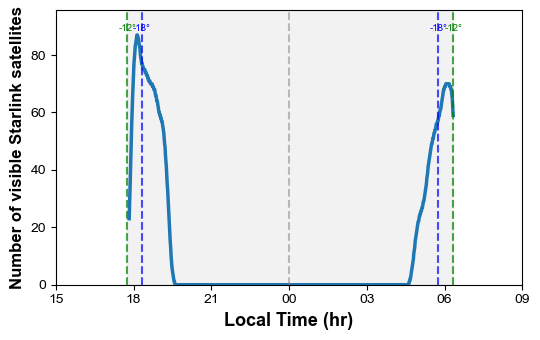}
\end{minipage}
\caption{Starlink satellites visible from the Xinglong Observatory in summer (Jun 15) (left) and winter (Dec 15) (right), versus time of night. Vertical lines indicate the times at which the sun reaches elevations $-12^\circ$ and $-18^\circ$ degrees for the ground observer.}
\label{fig:2}
\end{figure}

 Figure~\ref{fig:1} shows the number of Starlink satellites in orbit, the number of visible Starlink satellites at the Xinglong Observatory, and their corresponding visibility ratio for each sampling date. The total number of in-orbit Starlink satellites exhibits an overall upward trend with successive launches. As the number of Starlink satellites \deleted{launched increased} \added{increases}, the number of visible Starlink satellites \deleted{also increased overall} \added{similarly follows an upward trajectory}. As shown in the plot, the fraction of visible satellites in summer is significantly higher than in other months. We then simulate the number of these Starlink satellites that are visible as a function of date and time of night. \deleted{As a representative case, the number of visible Starlink satellites was simulated both in summer and winter.} \added{As a representative case, we simulate the visibility of Starlink satellites during both the summer and winter.} The number of visible Starlink satellites in summer and winter is presented in Figure~\ref{fig:2}, respectively. \deleted{It can be seen that the visibility duration of the satellites is longer in summer} \added{Notably, the visibility window is significantly protracted during summer}. During this \deleted{season} \added{period}, satellites \deleted{are visible all night} \added{remain illuminated throughout the night}, \deleted{from the end of evening nautical twilight until the beginning of morning nautical twilight} \added{spanning the entire interval between evening and morning nautical twilights}. In contrast, \added{during} \deleted{in} winter, satellites are primarily visible only during twilight.
 
\section{Detecting artificial satellite trails}
\label{sect:trails}
To assess the impact of artificial satellites on observations with the 80-cm telescope and the MST, \added{we} \deleted{We applied} \added{apply} the Probabilistic Hough Transform (PHT) to detect satellite trails in \deleted{images taken by the 80-cm telescope in 2019 and 2023, and MST in 2023} \added{the 80-cm telescope (2019, 2023) and MST (2023) datasets}. \deleted{We then calculated} \added{We subsequently calculate} the fraction of \deleted{trail-containing} \added{trail-contaminated} images \deleted{for each month from January to December} \added{on a monthly basis}. \added{To account for the non-uniform observation time across different months, we further determine the hourly trail-contaminated fraction for both facilities. Furthermore, we analyze the monthly fraction of images containing satellite trails
 categorized by solar elevation angle intervals, specifically focusing on the ranges from $-20^{\circ}$ to $-12^{\circ}$
and$-35^{\circ}$ to $-25^{\circ}$.}

\added{Although individual exposure times vary (predominantly ranging from $100\ \mathrm{s}$ to $300\ \mathrm{s}$), we have verified that the distribution of these durations is statistically consistent across different months. Since a satellite typically crosses the field of view in much less than $100\ \mathrm{s}$, any exposure within this range is equally capable of capturing a trail. Thus, the ``fraction of contaminated images'' serves as a robust proxy for the probability of interference per observing session.}

\subsection{Data}
We \deleted{utilized} \added{utilize} a total of 96,450 images captured by the 80-cm telescope at the Xinglong Observatory in 2019 and 2023, \deleted{respectively,} along with 27,205 images obtained by the MST in 2023. \deleted{The satellite trails within these images were detected} \added{We detect satellite trails within these datasets} using the PHT. Due to telescope maintenance \deleted{during} \added{in} July and August, no observations \deleted{were made} \added{acquired} \deleted{in} \added{during} these two months. The 80-cm telescope, known as the Tsinghua-NAOC Telescope (TNT), is a collaborative facility between Tsinghua University and NAOC. This instrument features an 80-cm aperture with a focal length of 8 m in a classical Cassegrain configuration, mounted on an equatorial platform. Its field of view spans $11.4' \times 11.1'$. All these parameters of instruments are summarized in Table~\ref{tab:80}. The MST project is a pathfinder of SiTian \citep{liu2021sitian,huang2025mini}. The MST is composed of three 30 cm catadioptric Schmidt telescopes: MST-001 (MST1), MST-002 (MST2) and MST-003 (MST3). One filter for each telescope, i, g and r band for MST1, MST2 and MST3, respectively, a wide $2.29^{\circ} \times 1.53^{\circ}$ field of view, and a 9K×6K scientific CMOS detector \citep{he2025mini,zhang2025mini}. All these parameters of instruments are summarized in Table~\ref{tab:Mini-SiTian}. 

\begin{table}[H]
\centering
\caption{Summary of the 80-cm Telescope}
\label{tab:80}
\begin{tabular}{>{\bfseries}l l}
\toprule
\textbf{Parameter Category} & \textbf{Specification} \\
\midrule
\multicolumn{2}{l}{\textbf{Telescope Parameters}} \\
\cmidrule(r){1-2}
Primary Mirror diameter 
& \SI{80}{\centi\meter} \\
Focal Ratio 
& f/3 \\
Corrected Focal Ratio 
& f/10 \\
Focal Plane Scale 
& \SI{25.8}{\arcsecond\per\milli\meter} \\
Optical System 
& Classical Cassegrain \\
Mount Type 
& Equatorial Mount \\
Field of View 
& $11.4' \times 11.1'$ \\
\addlinespace[0.2cm]
\multicolumn{2}{l}{\textbf{Camera Parameters}} \\
\cmidrule(r){1-2}
Model 
& PI VersArray 1300B LN Back-Illuminated Scientific CCD \\
Pixel Array 
& $1,340 \times 1,300$ pixels \\
Pixel Size 
& \SI{20}{\micro\meter} \\
Pixel Scale 
& \SI{0.52}{\arcsecond\per\pixel} \\
Operating Temperature 
& \SI{-110}{\degreeCelsius} \\
\bottomrule
\end{tabular}
\vspace{0.2cm}
\end{table}

\begin{table}[H]
\centering
\caption{Summary of the MST Telescope Array }
\label{tab:Mini-SiTian}
\begin{tabular}{>{\bfseries}l l}
\toprule
\textbf{Parameter Category} & \textbf{Specification} \\
\midrule
\multicolumn{2}{l}{\textbf{Telescope Parameters}} \\
\cmidrule(r){1-2}
Primary Mirror diameter 
& \SI{30}{\centi\meter} \\
Number of telescopes
& 3 \\
Focal Ratio 
& f/3 \\
Optical System 
& Catadioptric Schmidt \\
Mount Type 
& Fork-type equatorial \\
Field of View 
& $5^\circ \times 5^\circ$ \\
\addlinespace[0.2cm]

\multicolumn{2}{l}{\textbf{Camera Parameters}} \\
\cmidrule(r){1-2}
Model 
& ZWO ASI6200MM Pro CMOS \\
Pixel Array 
& 9,576 $\times$ 6,388 pixels \\
Pixel Size 
& \SI{3.76}{\micro\meter} \\
Pixel Scale 
& \SI{0.862}{\arcsecond\per\pixel} \\
Camera FOV 
& $2.29^\circ \times 1.53^\circ$ \\
Operating Temperature 
& Cooling \SI{30}{\degreeCelsius} below ambient \\
\bottomrule
\end{tabular}

\vspace{0.2cm}
\end{table}

\subsection{Method}
This study \deleted{employs} \added{utilizes} the PHT to detect satellite trails. The Hough line detection algorithm \deleted{operates by transforming} \added{transforms} the problem of line detection in image space into \deleted{the problem of} peak detection \deleted{in} \added{within a} parameter space.\deleted{By identifying peaks within the parameter space, the algorithm effectively accomplishes the task of line detection, enabling the robust detection of linear patterns such as satellite trails.} \added{This approach enables the robust identification of linear features, such as satellite trails, by isolating these local maxima.} The parameter space \deleted{utilizes} \added{is defined by} a polar coordinate system, \added{where} any line in the image \deleted{can be} \added{is} represented as a point \added{governed} \deleted{defined} by the equation $r = x \cos \theta + y \sin \theta$, where $r$ is the perpendicular distance from the origin to the line, and $\theta$ is the angle of the normal with the horizontal axis \citep{stoppa2024automated}. Each point \((r, \theta)\) in the parameter space \deleted{corresponds} \added{maps} to a unique line in the image space. Conversely, a single point in the image space \deleted{is represented as} \added{generates} a sinusoidal curve in the parameter space. For every candidate pixel \deleted{potentially belonging to} \added{associated with} a satellite trail, the \added{algorithm} \deleted{Hough Transform algorithm evaluates} \added{identifies} all potential lines passing through that pixel, each defined by a distinct \((r, \theta)\) pair. \deleted{Through this transformation, each image-space pixel generates a sinusoidal curve in the parameter space.} The intersection points of these curves, derived from distinct pixels, signify a collective agreement on the \deleted{existence} \added{presence} of a \deleted{corresponding} straight line in the \deleted{image space} \added{original image}.

\begin{figure}[H]
\centering
\begin{minipage}{0.45\textwidth}
\centering
\includegraphics[width=\linewidth]{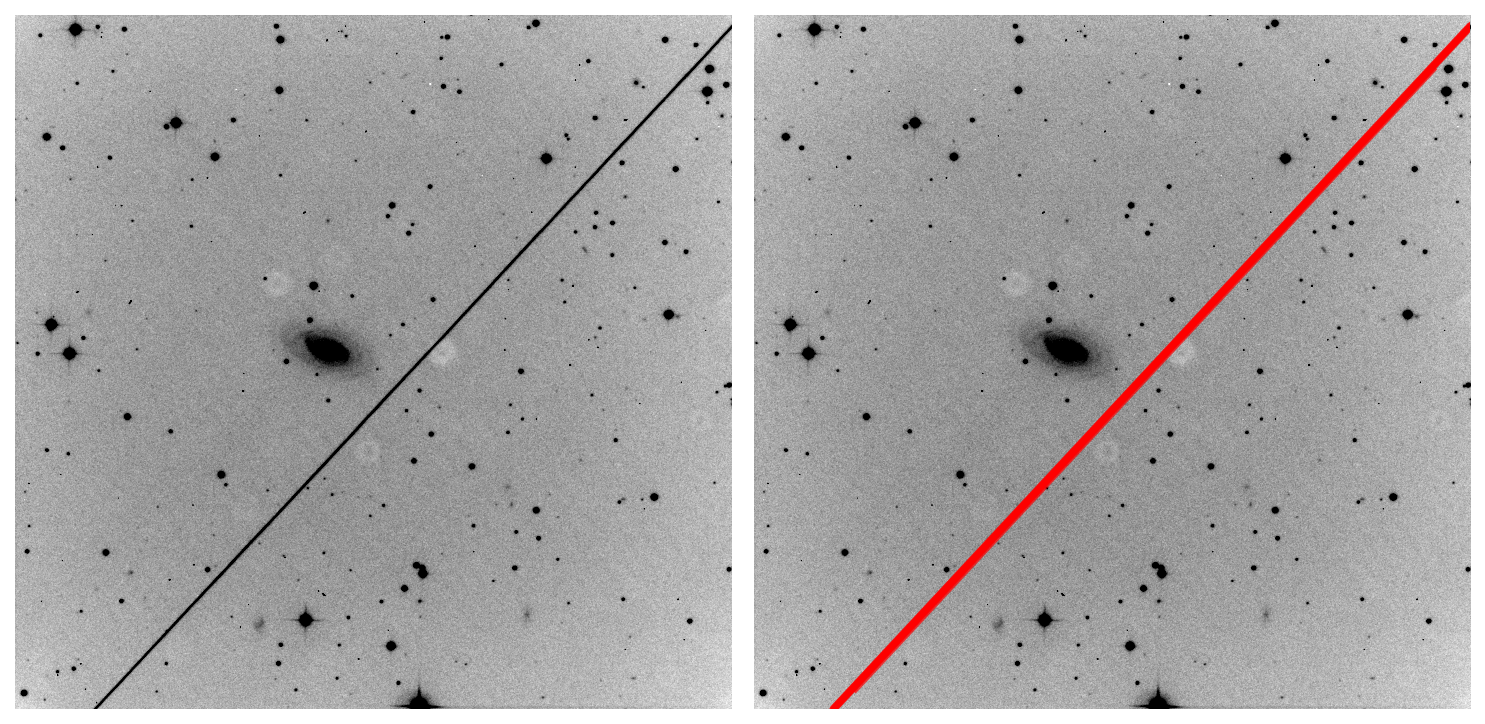}
\end{minipage}
\hfill
\begin{minipage}{0.5\textwidth}
\centering
\includegraphics[width=\linewidth]{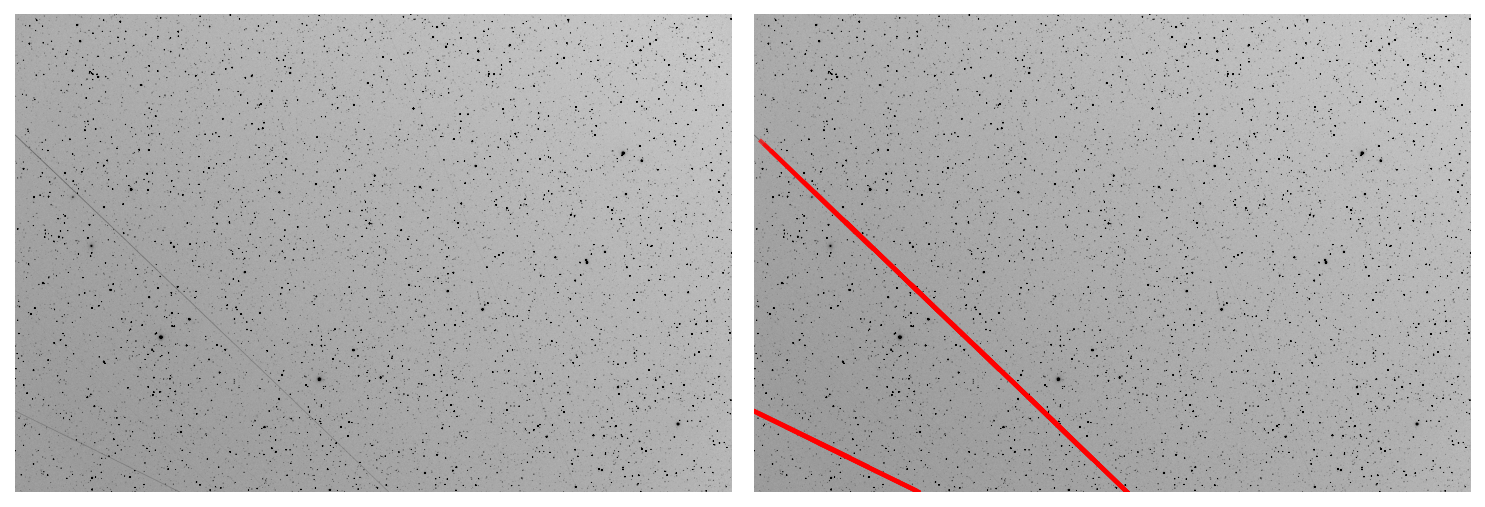}
\end{minipage}
\caption{Original image with satellite trails detected by the PHT and marked by red lines, the 80-cm telescope (left) and the MST2 (right).}
\label{fig:3}
\end{figure}

The PHT \deleted{is} \added{serves as} a computationally efficient variant of the standard Hough Transform, \deleted{which detects lines} \added{detecting linear features} by processing \added{random} subsets of image points. \deleted{Algorithm parameters were rigorously optimized} \added{We rigorously optimized the algorithm parameters as follows}:
\begin{itemize}    
\item \textit{Distance resolution}
: 3~pixels (\added{balancing} computational speed and \added{spatial} precision)    
\item \textit{Angular resolution}: $1^\circ$ (providing full $180^\circ$ coverage)    
\item \textit{Min line length}
: 1000~pixels (aligned with \added{typical} satellite trail characteristics)    
\item \textit{Max gap}
: 300~pixels (robust to occlusion effects)    
\item \textit{Accumulator threshold}: \added{A} Critical value separating \textit{bona fide} 
trails from stochastic noise
\end{itemize}
\deleted{The parameter set comprehensively accounts for the geometric characteristics of LEO satellites.} \added{This configuration effectively accounts for the specific geometric profiles of LEO satellite streaks.} Figure~\ref{fig:3} \deleted{display} \added{displays} images containing satellite trails \deleted{identified using the PHT, captured by} \added{detected via the PHT from} the 80-cm telescope and MST2, respectively.

\subsection{Results}
\begin{figure}[H]    
\centering    
\includegraphics[width=0.7\linewidth]{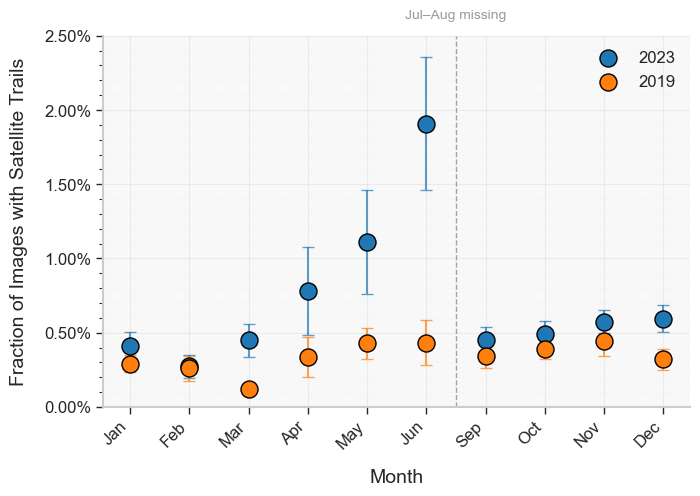}    
\caption{Fraction of images with satellite trails (80-cm telescope, 2019 vs. 2023): orange dots for 2019, blue dots for 2023.}    
\label{fig:impact_80}
\end{figure}

\begin{figure}[H]    
\centering    
\includegraphics[width=0.7\linewidth]{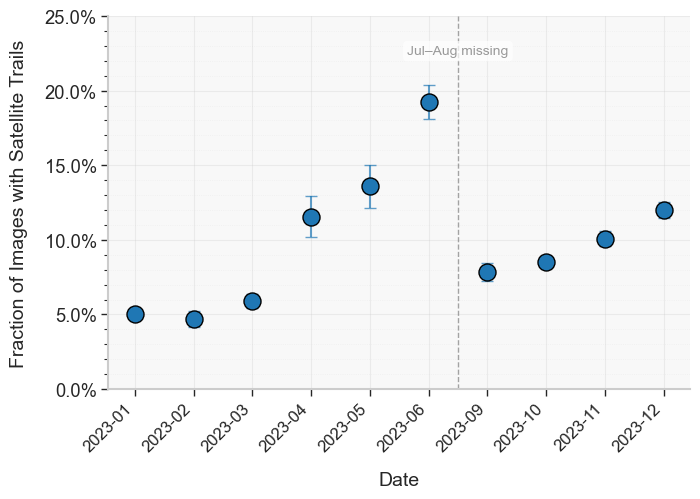}    
\caption{Fraction of images with satellite trails \deleted{captured} \added{acquired} by MST2 in 2023.}    
\label{fig:impact_mst}
\end{figure}

Figure~\ref{fig:impact_80} shows the fraction of images containing satellite trails \deleted{captured by} \added{acquired by} the 80-cm telescope in 2019 and 2023, \deleted{respectively,} while Figure~\ref{fig:impact_mst} shows this fraction for MST2 observations in 2023. \added{An} \deleted{It is evident that there is an} overall upward trend in \added{image contamination} \deleted{the fraction of images containing satellite trails} \added{is evident,} correlating with \deleted{increasing satellite deployments} \added{expanded satellite constellations}. For the 80-cm telescope, the \added{contamination} fraction \deleted{increased} \added{rises} from an average of 0.34\% in 2019 to 0.7\% in 2023. Due to its \deleted{larger} \added{wider} field of view \added{(FoV)}, the MST2 \deleted{has} \added{exhibits} a markedly higher fraction of images with satellite trails than the 80-cm telescope. \deleted{The} \added{Specifically, the} fraction of images containing satellite trails \deleted{increased} \added{grows} from 5\% in January 2023 to 12\% in December 2023, reaching a peak of 19\% in summer. 

%%%%%%%%%%%%%%%%%%%%%%%%%%%%%%%%%%%%%%%%%%%%%%%%%%%%%%%%%%%%%%
%%     Examples for figures using graphicx for LaTeX 2e
%%               -- our recommended way for embodying graphics
%%%%%%%%%%%%%%%%%%%%%%%%%%%%%%%%%%%%%%%%%%%%%%%%%%%%%%%%%%%%%%
%
%      A figure as large as the width of the column
%-------------------------------------------------------------
  
%
%      One column rotated figure
%-------------------------------------------------------------

\begin{figure}[H]    
\centering    
\includegraphics[width=0.8\linewidth]{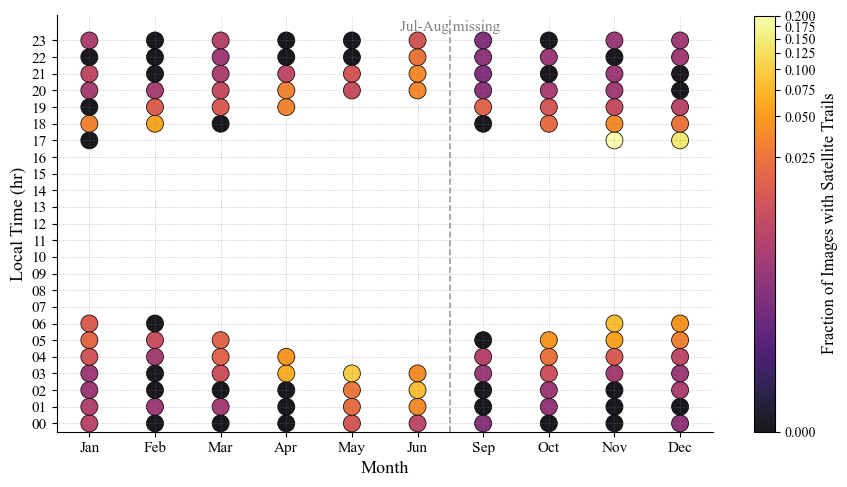}    
\caption{Hourly fraction of images with satellite trails acquired by the 80-cm telescope in 2023.}    
\label{fig:80_hour}
\end{figure}

\begin{figure}[H]    
\centering    
\includegraphics[width=0.8\linewidth]{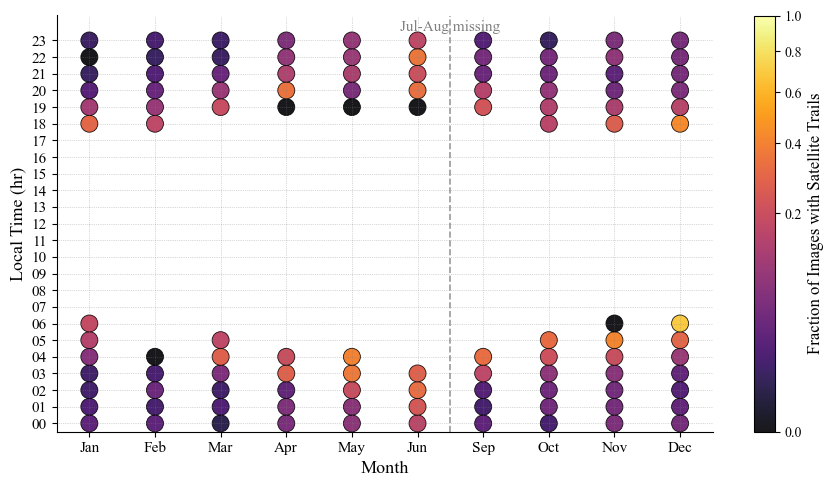}    
\caption{Hourly fraction of images with satellite trails acquired by the MST2 in 2023.}    
\label{fig:mst_hour}
\end{figure}

\begin{figure}[H]    
\centering    
\includegraphics[width=0.8\linewidth]{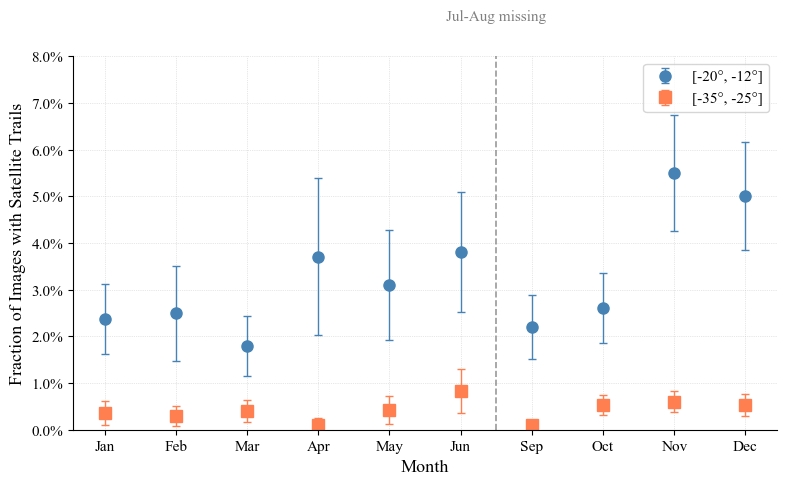}    
\caption{Fraction of images with satellite trails acquired by the 80-cm telescope in 2023. Blue dots and orange squares represent images taken within solar elevation ranges of $-20^{\circ}$ to $-12^{\circ}$ and $-35^{\circ}$ to $-25^{\circ}$, respectively.}    
\label{fig:80_sun}
\end{figure}

\begin{figure}[H]    
\centering    
\includegraphics[width=0.8\linewidth]{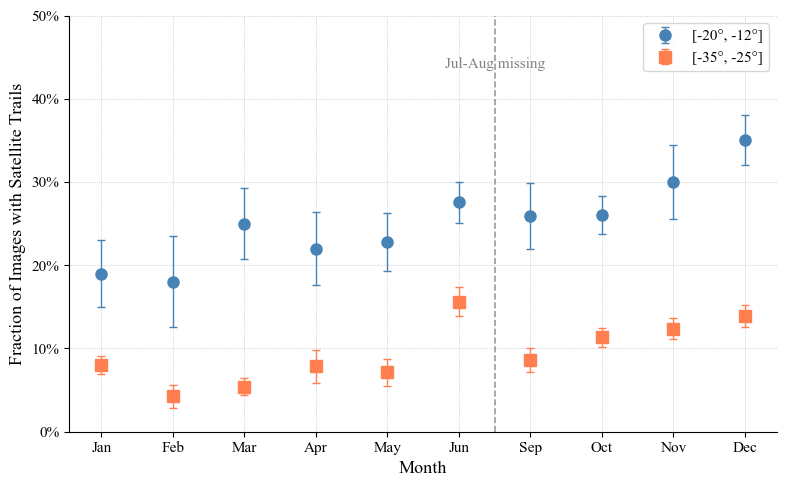}    
\caption{Fraction of images with satellite trails acquired by the MST2 in 2023. Blue dots and orange squares represent images taken within solar elevation ranges of $-20^{\circ}$ to $-12^{\circ}$ and $-35^{\circ}$ to $-25^{\circ}$, respectively.}    
\label{fig:mst_sun}
\end{figure}

%----------------------------------------------------- Figs 3 & 4:
\added{To account for the non-uniform observation time across different months, Figure~\ref{fig:80_hour} and Figure~\ref{fig:mst_hour} show the monthly distribution of the hourly trail-contaminated fraction for the 80-cm telescope and MST2 throughout 2023, respectively. Furthermore, Figure~\ref{fig:80_sun} and Figure~\ref{fig:mst_sun} illustrate the monthly fraction of images containing satellite trails categorized by solar elevation angle intervals, specifically for the ranges from $-20^{\circ}$ to $-12^{\circ}$
and$-35^{\circ}$ to $-25^{\circ}$, for both facilities.}
\added{As illustrated, with the increasing number of satellite deployments, the fraction of trail-contaminated images exhibits an overall upward trend. The impact is particularly severe during twilight; for the wide-field MST2, the affected fraction rose from less than 20\% in January to approximately 35\% by December 2023. We note that certain short-term fluctuations, such as the increase between March and April, are associated with larger statistical uncertainties due to smaller sample sizes in specific months (e.g., $N_{\text{Apr}}=606$ vs. $N_{\text{Mar}}=2491$), as reflected by the provided error bars.}

\added{Regarding deep-night observations, the contamination fraction during the summer is notably higher than in winter. This trend is a direct physical consequence of the LEO orbital geometry during the summer solstice. At the observatory's latitude, the sun-illumination window for LEO satellites is significantly extended during the summer, persisting even through the midnight hours.}

As expected by \citet{mcdowell2020low} , \added{twilight and} summer observations are particularly vulnerable to \deleted{contamination from} satellite \added{trail contamination}. \added{For deep-night observations, the fraction of contaminated images during summer is notably higher than in other months.} Telescopes with \deleted{a large field of view} \added{wider FoVs} are significantly more affected. Specifically, approximately 19\% of the images \deleted{captured} \added{acquired} by the MST2 during the summer \deleted{were contaminated by satellite trails} \added{exhibit satellite trails}. \added{Furthermore, while less than 20\% of twilight images were affected in January 2023, this fraction increased to approximately 35\% by December 2023.} As of December 31, 2023, SpaceX \deleted{had launched} \added{has deployed} more than 5,000 satellites cumulatively. \deleted{So far, although} \added{While} the fraction of \added{contaminated} images \deleted{containing satellite trails is increasing} \added{grows} with \deleted{the growing number of satellite deployments} \added{constellation expansion}, the scientific operations of both the 80-cm telescope and the MST have not yet been severely compromised. However, the number of satellites and space debris will only increase in the future, \deleted{and} SpaceX has \deleted{been granted approval by} \added{received} FCC \added{approval} to operate up to 42,000 satellites in total \citep{halferty2022photometric}, \added{and} the proliferation of artificial satellites is forecast to \deleted{escalate beyond} \added{exceed} 100,000 active units in \deleted{near-Earth orbits} \added{LEO} before 2030 \citep{krantz2021characterizing}. We estimate that if the full constellation of 42,000 Starlink satellites is deployed, nearly all MST images \deleted{captured} \added{acquired} during \deleted{summer} \added{twilight}  \deleted{would} \added{will} be affected. \added{Furthermore,} \deleted{Once} \added{should} the constellation scale \deleted{reaches} \added{reach} 100,000 satellites, virtually all MST images \added{acquired in summer} \deleted{would} \added{will} be affected. \deleted{regardless of the season}

\section{Aperture photometry }
\label{sect:photometry}
To assess the impact of satellite trails on sources, we \deleted{quantified} \added{quantify} the fraction of affected sources in the \deleted{previously detected trail images from both} \added{detected trail-contaminated images for} the 80-cm telescope and the MST2. In addition, \added{we perform} aperture photometry \deleted{was performed} specifically on the MST2 images affected by satellite trails \deleted{ to obtain the magnitudes of the sources} \added{to derive source magnitudes}. Subsequently, by cross-matching with the Gaia catalog, we \deleted{calculated} \added{calculate} the residuals between the measured magnitudes and their Gaia counterparts: \added{$\Delta \mathrm{mag} =$} $\mathrm{mag}_{\mathrm{measure}} - \mathrm{mag}_{\mathrm{Gaia}}$. Finally, using \deleted{the $\mathrm{mag}_{\mathrm{measure}} - \mathrm{mag}_{\mathrm{Gaia}}$ values} \added{$\Delta \mathrm{mag}$} of sources located near the trails as a sample, we \deleted{calculated} \added{calculate} the distribution of this sample relative to \deleted{the $\mathrm{mag}_{\mathrm{measure}} - \mathrm{mag}_{\mathrm{Gaia}}$ values} \added{$\Delta \mathrm{mag}$} of the entire source population. \deleted{From these data, we analyzed} \added{These data allow us to analyze} the impact of satellite trails on sources at various distances and across a range of \deleted{brightness levels } \added{magnitudes}.

\subsection{Method}
\added{To analyze} images containing satellite trails, \deleted{the coordinates of the trails were first located and their width was determined, followed by measuring the distance from sources to these trails} \added{we first determine the trail coordinates and widths, subsequently measuring the distance between each source and the trails}. Based on the results \deleted{from} \added{of} our subsequent analysis, we \deleted{selected} \added{adopt} a distance of 12 pixels as the threshold\added{;}\deleted{, any} sources within this distance \deleted{from a trail were} \added{are} considered \added{to be} affected. We then \deleted{quantified} \added{quantify} the fraction of affected sources\deleted{, respectively,} in \deleted{images with satellite trails captured from January to December 2023 by} \added{the 2023 (January--December) trail-contaminated datasets for} the 80-cm telescope and the MST2 \added{respectively}.

Furthermore, \deleted{Aperture photometry was performed} \added{we perform aperture photometry} on the \added{trail-contaminated} MST2 images. \deleted{containing satellite trails to obtain source magnitudes} \added{To ensure the highest photometric precision across varying atmospheric conditions, we employed a set of fixed circular apertures with diameters ranging from 6 to 16 pixels
($5.2 ^{\prime\prime}$ to $13.8 ^{\prime\prime}$). For each image, the optimal aperture size was dynamically selected by minimizing the photometric uncertainty and maximizing the signal-to-noise ratio (SNR) based on the local seeing (FWHM). The majority of the samples utilized optimal apertures between 8 and 12 pixels ($6.9 ^{\prime\prime}$ to $10.3 ^{\prime\prime}$).} 

The \deleted{results were then cross-matched} \added{resulting magnitudes are cross-matched} with the Gaia catalog for \deleted{calibration} \added{photometric calibration}. We \deleted{then calculated} \added{subsequently calculate} the residuals between our measured magnitudes and the corresponding magnitudes from the Gaia catalog: \added{$\Delta \mathrm{mag} =$} $\mathrm{mag}_{\mathrm{measure}} - \mathrm{mag}_{\mathrm{Gaia}}$. All sources \deleted{were binned in intervals of 0.5 magnitudes} \added{are binned in 0.5-mag intervals} based on their Gaia magnitudes. \deleted{The standard deviation $\sigma_i$ was calculated} \added{We calculate the standard deviation $\sigma_i$} for each magnitude bin \deleted{which was then used to compute the standardized residual $\sigma$ for each source,} \added{ to derive the standardized residual $\sigma$ for each source} :  $\sigma = (\mathrm{mag}_{\mathrm{measure}} -\mathrm{mag}_{\mathrm{Gaia}})/\sigma_i$. The sources \deleted{were} \added{are} grouped separately based on their distance from the satellite trail \added{(in 0.5-pixel steps)} and their magnitude \added{(in 0.5-mag steps)}. \deleted{The median standardized residuals were then calculated separately for different distance bins and different magnitude bins, thus assessing} \added{By computing the median standardized residuals for each bin, we assess} the impact of satellite trails on sources at various distances and with different \deleted{brightness levels } \added{magnitudes}.

\subsection{Results}
\begin{figure}[h]
\centering
\begin{minipage}{0.47\textwidth}
\centering
\includegraphics[width=\linewidth]{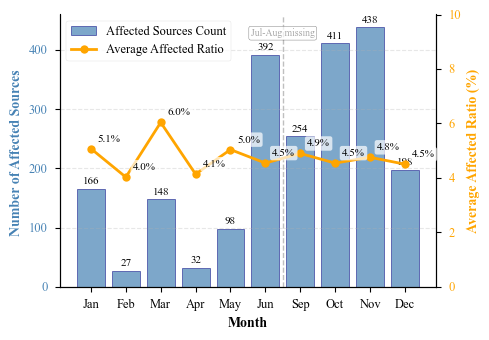}
\end{minipage}
\hfill
\begin{minipage}{0.47\textwidth}
\centering
\includegraphics[width=\linewidth]{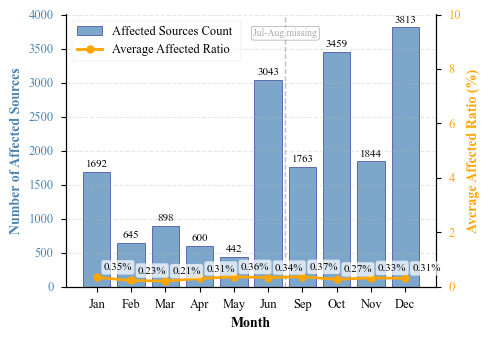}
\end{minipage}
\caption{Monthly average of the \deleted{number} \added{count} of affected sources and their fraction per \deleted{satellite-trail} \added{trail-contaminated} image, shown for the 80-cm telescope (left ) and the MST2 (right ), from January – December 2023. Blue bars indicate the counts, and the orange line represents the fraction.}
\label{fig:6}
\end{figure}

Figure~\ref{fig:6} shows the monthly averaged \deleted{number} \added{count} of affected sources and their fraction per \deleted{satellite-trail} \added{trail-contaminated} image \deleted{, captured by}  \added{for} the 80-cm telescope and the MST2 throughout 2023. The values \deleted{were} \added{are} first calculated for each trail-contaminated image and then averaged monthly. For the 80-cm telescope, the average fraction of affected sources per trail-contaminated image is approximately 5\%, with minimal monthly fluctuation. In contrast, the fraction for the MST2 Array \deleted{was} \added{is} approximately 0.3\%, also showing little variation from month to month.

\begin{figure}[H]    
\centering    
\includegraphics[width=0.9\linewidth]{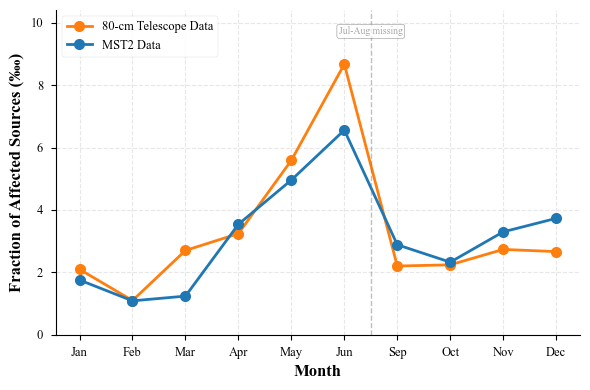}    
\caption{Fraction of affected sources in images \deleted{captured} \added{acquired} by the 80-cm telescope and the MST2 from January to December 2023: Orange represents the 80-cm telescope, and blue represents the MST2.}    
\label{fig:7}
\end{figure}

Subsequently, by incorporating the monthly fraction of images containing satellite trails, we \deleted{calculated} \added{calculate} the overall fraction of sources affected with images for both the 80-cm telescope and the MST2. Figure~\ref{fig:7} shows the fraction of affected sources in images \deleted{captured} \added{acquired} by the 80-cm telescope and the MST2 from January to December 2023. \deleted{It can be seen that the fraction of affected sources shows little difference between the images from the 80-cm telescope and the MST2} \added{The data reveal a negligible discrepancy in the affected source fraction between the two facilities}. This fraction, while currently \deleted{small} \added{modest}, \deleted{is increasing with the growing number of satellite deployments} \added{exhibits an upward trend correlated with the expanding satellite constellations}. In particular, a pronounced peak \deleted{is evident in the affected fraction} \added{emerges} during summer.

\begin{figure}[H]    
\centering    
\includegraphics[width=0.8\linewidth]{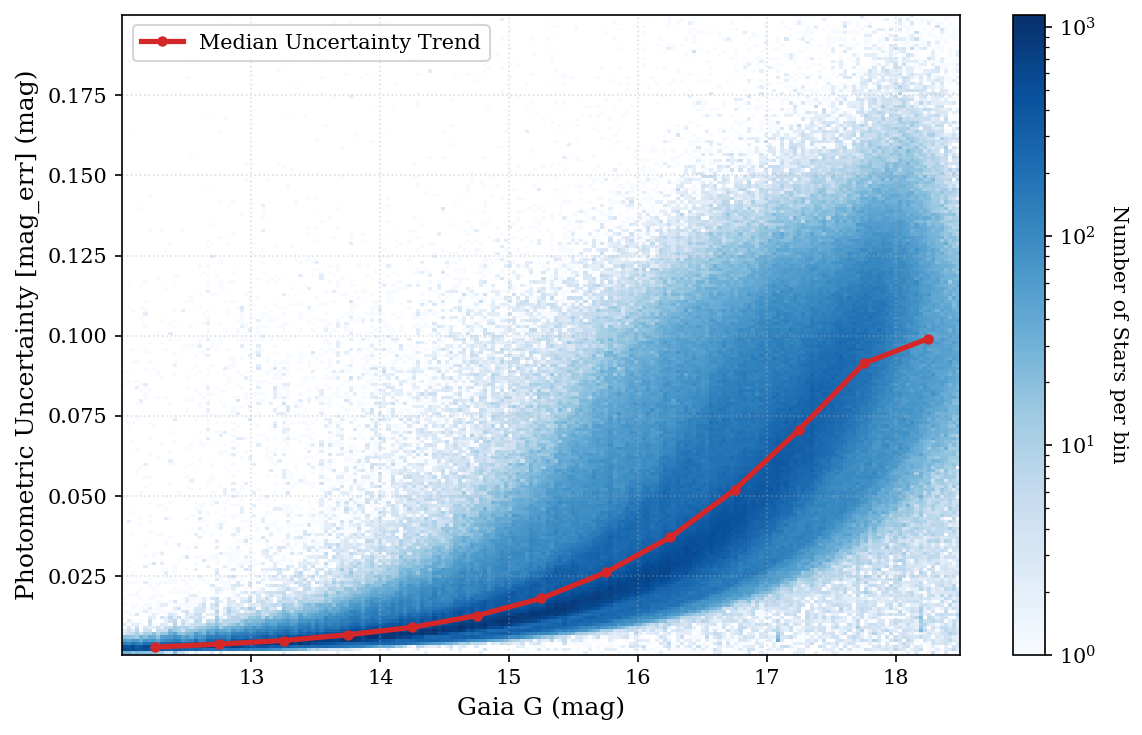}    
\caption{2D density distribution of photometric uncertainties ($\mathrm{mag\_err}$) as a function of Gaia G magnitude for the MST2 observation system.}    
\label{fig:mag_err}
\end{figure}

\added{To evaluate the performance of the MST2 photometric system, we analyzed the relationship between typical photometric uncertainties and Gaia G magnitudes. As illustrated in Figure~\ref{fig:mag_err}, the median uncertainty trend (red line) clearly reveals the evolution of photon noise as constrained by detector performance. The photometric error remains remarkably low ($< 0.02$ mag) for $G < 15$, rising to approximately $0.1$ mag at $G \approx 18.2$. Since all satellite trail interferences are quantified using the standardized residual $\sigma$ (where residuals are normalized by the intrinsic uncertainty $\sigma_i$ at each magnitude level), a consistent assessment of the trail impact is maintained across varying signal-to-noise ratios.}

\begin{figure}[H]
\centering
\begin{minipage}{0.33\textwidth}
\centering
\includegraphics[width=\linewidth]{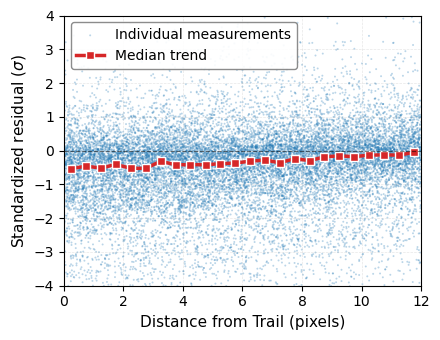}
\end{minipage}
\hfill
\begin{minipage}{0.33\textwidth}
\centering
\includegraphics[width=\linewidth]{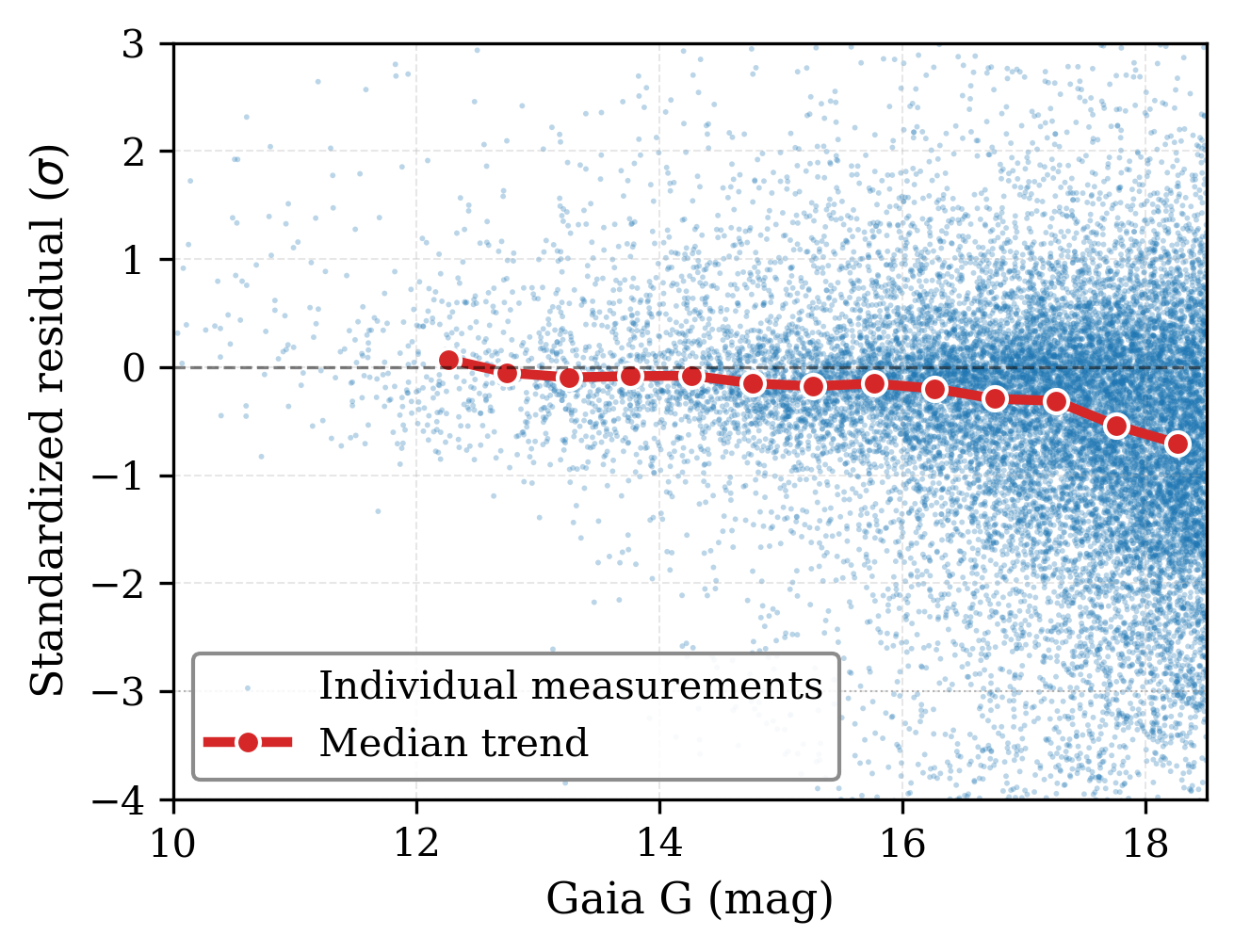}
\end{minipage}
\hfill
\begin{minipage}{0.33\textwidth}
\centering
\includegraphics[width=\linewidth]{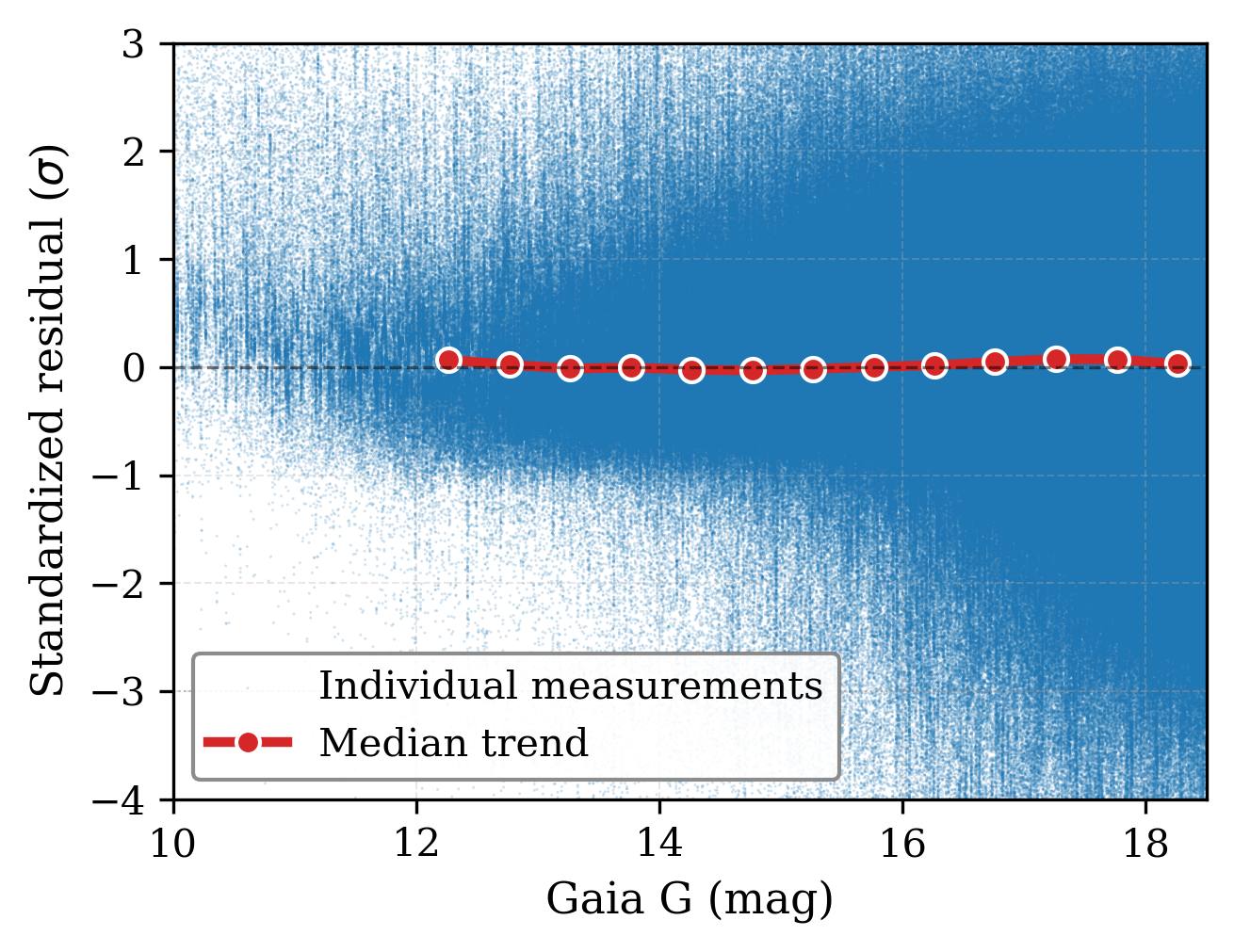} % 请替换为第三张图的文件名
\end{minipage}
\caption{Plots of the standardized residual $\sigma$ versus : (left) the distance from the satellite trail for sources within 12 pixels, (center) the magnitude for sources within 12 pixels, and (right) the magnitude for all sources. The blue dots represent the measured values for each individual source, the red trend lines represent the median standardized residual values corresponding to different distance bins and different magnitude bins, respectively.}
\label{fig:8}
\end{figure}

Figure~\ref{fig:8} separately shows the relationships between  $\sigma$ and the distance from the satellite trail, \added{as well as} between $\sigma$ and \deleted{the} magnitude. The \added{left} panel shows the $\sigma$ of sources located within 12 pixels of the satellite trail as a function of their distance from the trail. The \added{middle} panel shows the $\sigma$ of these same sources (within 12 pixels) as a function of their magnitude. \added{For comparison}, the \added{right} panel shows $\sigma$ for all sources as a function of their magnitude. The red trend lines represent the median standardized residual values \deleted{corresponding to} \added{calculated for} different distance bins and different magnitude bins, respectively.

The results indicate that the impact \added{of satellite trails} intensifies with \deleted{proximity to the satellite trail} \added{decreasing distance to the trail}. Specifically, the closer a source is to the satellite trail, the more negative the median standardized residual ($\sigma$) becomes. The deviation \deleted{increased} \added{increases} from approximately 0.1$\sigma$ at a distance of 11 pixels from the trail to approximately 0.5$\sigma$ at a distance of 1 pixel. This trend indicates that sources appear progressively brighter in our measurements relative to the Gaia reference magnitude \deleted{the closer they are to the satellite trail} \added{as they approach the trail}. At a distance of 12 pixels, the median standardized residual approaches zero, suggesting that sources beyond this threshold are virtually unaffected. \deleted{With respect to magnitude} \added{Regarding brightness}, fainter sources experience a \added{more pronounced} impact. For sources \deleted{with magnitudes} fainter than 16 mag, the median standardized residual \deleted{progressively} shifts toward negative values, the deviation \deleted{increased} \added{increases} from approximately $0.15\sigma$ at 16 \added{mag} to approximately $0.7\sigma$ at 18 \added{mag}.

\added{To further investigate the modulation of photometric interference by atmospheric seeing, we performed a sub-group analysis based on the seeing conditions of the images: Good ($<$ $3.5^{\prime\prime}$
 ), Median ($3.5^{\prime\prime}$-$4.5^{\prime\prime}$), and Poor ($>$ $4.5^{\prime\prime}$).}

\begin{figure}[H]    
\centering    
\includegraphics[width=0.8\linewidth]{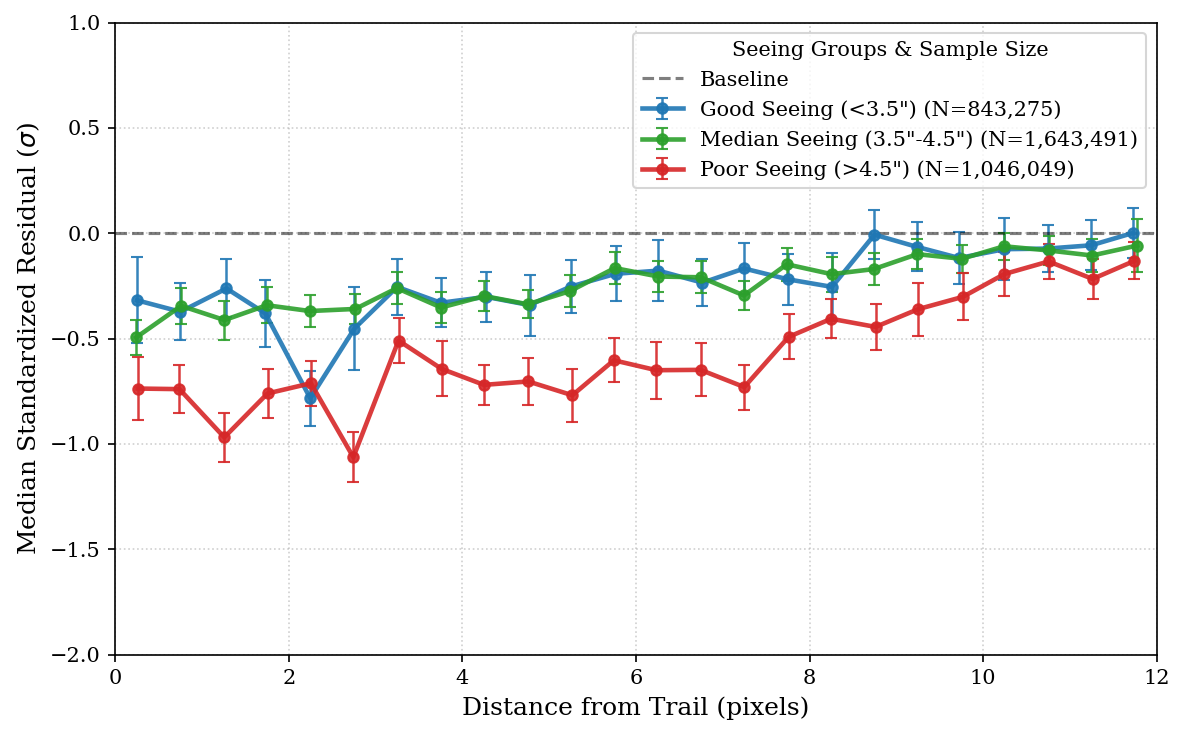}    
\caption{Median standardized photometric residuals ($\sigma$) as a function of the distance from the satellite trail, categorized by atmospheric seeing conditions.}    
\label{fig:Seeing Impact}
\end{figure}

\added{As shown in Figure~\ref{fig:Seeing Impact}, the median standardized residuals ($\sigma$) exhibit a pronounced dependence on seeing. In images with Poor Seeing ($>$ $4.5^{\prime\prime}$), the satellite trails present a broader transverse profile, which significantly increases the distance threshold at which scattered light enters the photometric aperture. Quantitatively, the median standardized residuals ($\sigma$) in the Poor Seeing group fail to return to the baseline even as the distance from the trail increases to 12 pixels. The magnitude of deviation in the Poor Seeing group is consistently and significantly larger than that in the Good Seeing group, while the latter shows much smaller deviations overall. Specifically, for the Poor Seeing group, the deviation of median standardized residuals remain greater than 0.5 $\sigma$ for all sources within 8 pixels of the trail, highlighting the extended spatial impact of satellite contamination under suboptimal atmospheric conditions.}

\section{Discussion}
\label{sect:discussion}
As \deleted{demonstrated by the} simulation results and observational data \deleted{presented above} \added{indicate}, the impact of artificial satellites on astronomical observations is \deleted{progressively increasing} \added{escalating} with the growing deployment of satellites. For wide-field telescopes such as the MST, the fraction of images containing satellite trails is significantly higher. Although the number of images containing satellite trails is gradually rising, the scientific operations of the 80-cm telescope and the MST at the Xinglong Observatory have not been severely compromised by trails at present. The population of LEO satellites is undergoing dramatic expansion, with Starlink currently holding predominance. As the first large-scale constellation to achieve massive deployment, Starlink will not remain \deleted{solitary} \added{alone}: 43 mega-constellation projects are now in active development \citep{zhi2024multicolour}. Should SpaceX's \deleted{pending} regulatory \added{requests} for 42,000 Starlink satellites \deleted{materialize} \added{be approved}, \added{and} coupled with the implementation of other proposed constellations, the proliferation of artificial satellites is forecast to escalate beyond 100,000 active units in LEO before 2030 \citep{krantz2021characterizing}. These satellites will act as reflective surfaces for sunlight; thus, the impact on ground-based astronomical observations is projected to be significant.

Sustained astronomical monitoring programs require uninterrupted stabilized data acquisition. Satellite \deleted{crossings} \added{passages} within instrument FOVs generate: stray light contamination, detector saturation, flux contamination, and changes in the image background. Such effects irreversibly compromise critical research domains like time-domain astronomy, all-sky surveys, deep-field imaging, spectroscopic observations, and surface source imaging. Facilities conducting wide-field surveys, such as the Vera C. Rubin Observatory, experience particularly severe consequences \citep{zhi2024multicolour}. 

The impacts of large constellations of LEO satellites on astronomical research \deleted{programs} and public experience are highly variable, \added{with estimates ranging} \deleted{anywhere} between negligible and extreme. This variation hinges on observational factors such as the specific scientific or practical objectives, the facility's etendue, the observation strategy's flexibility for satellite avoidance, and the effectiveness of trail masking or removal in data processing. Furthermore, the severity of impact is critically dependent on the characteristics of the satellites: their total population, orbital \deleted{height} \added{altitude}, apparent \deleted{luminosity} \added{brightness}, \deleted{and} orientation, and the precision of \deleted{their} orbital ephemerides \citep{walker2020impact}.

Existing research has characterized such pixel contamination rates, with the current consensus indicating minimal overall pixel loss fractions. For example, extrapolation from current observations indicates that \added{a constellation of} 42,000 Starlink satellites would contaminate approximately 0.04\% of pixels in ZTF data over a year \citep{ref2}. Based on \deleted{current observational data} \added{our observations}, we \deleted{have also estimated} \added{estimate} the pixel loss fractions due to satellite trail contamination for the 80-cm telescope and the MST2. For the 80-cm telescope, 42,000 Starlink satellites would contaminate approximately 0.02\% of \deleted{its} pixels annually. For the MST2, the same number of satellites would \deleted{lead to contamination in} \added{affect} about 0.07\% of \deleted{its} pixels per year. Nevertheless, these impacts deserve serious consideration. Primarily, their removal necessitates significant computational expenditure. More critically, such streaks can obscure transient \added{events} or Solar system objects, potentially compromising discovery opportunities.

\begin{figure}[H]    
\centering    
\includegraphics[width=0.8\linewidth]{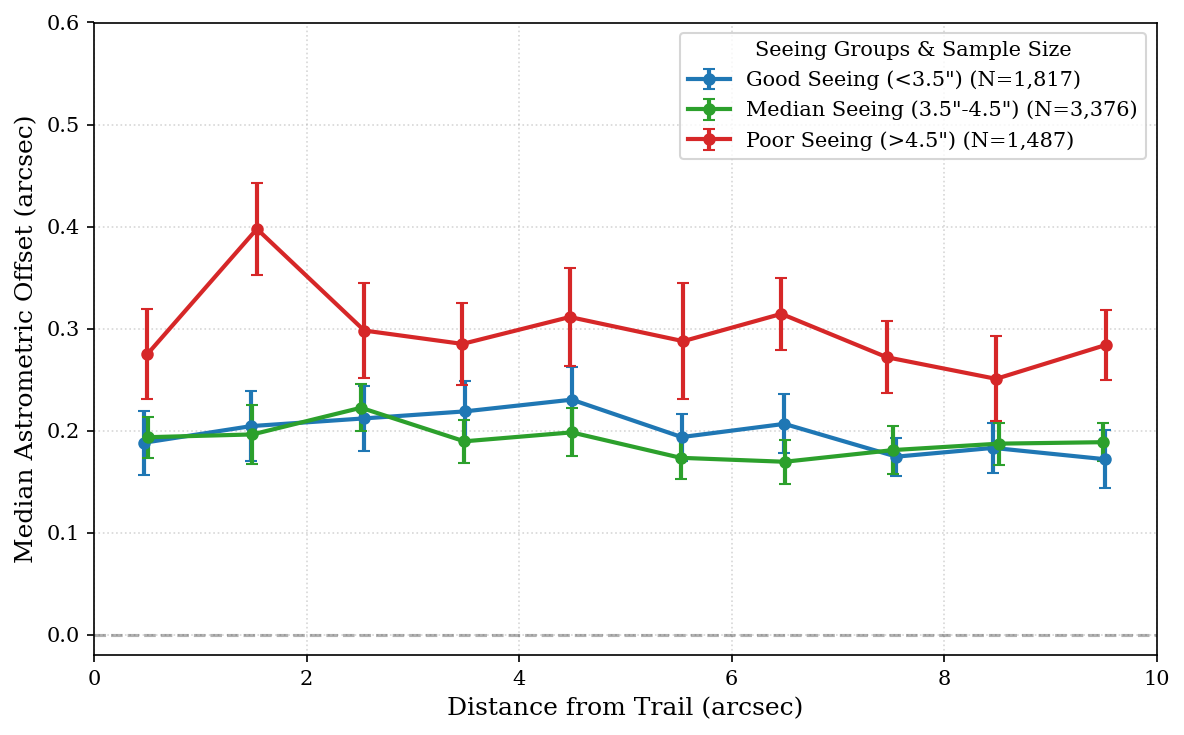}    
\caption{Median astrometric offsets between measured centroids and Gaia reference positions relative to the distance from the satellite trail.}   
\label{fig:Astrometric Precision}
\end{figure}

\added{We also discussed the impact of satellite trails on astrometric precision. Figure~\ref{fig:Astrometric Precision} illustrates the relationship between astrometric offset, cross-track distance, and seeing conditions. The results indicate that atmospheric seeing is the dominant factor governing astrometric uncertainty, significantly outweighing the direct contribution from the satellite trail itself. Even within the contaminated regions, the baseline seeing level (expressed as PSF dispersion) accounts for the majority of the positional uncertainty. The satellite trail acts only as a secondary factor under poor seeing conditions, inducing minor additional offsets at close range by disrupting the local symmetry of the background profile.}

SpaceX's next-generation satellite platform (Gen2 v2.0) is currently in development. The \deleted{pathfinder of Gen2} \added{precursor to this generation}, designated \added{as the} V2 Mini, has a surface area more than four times as great as the Gen 1 satellites \citep{mallama2023starlink}. Given the established correlation between spacecraft surface area and optical brightness, astronomers \deleted{expressed} \added{have voiced} significant concern about the brightness of \deleted{the} Starlink Gen2 satellites. To address the increased size of its satellites, SpaceX \deleted{implemented} \added{has employed} a \deleted{rigorous} \added{comprehensive} brightness reduction strategy. The most recent observations indicate that \deleted{the Gen2 Mini satellite is} \added{V2 Mini satellites are} actually fainter than the Gen 1 spacecraft \citep{mallama2023starlink}. The company’s brightness reduction strategies will face increased challenges with the deployment of future full-scale satellites \citep{mallama2023assessment}.

\section{Conclusions}
\label{sect:conclusion}
Based on simulation and observational data, it is evident that the \deleted{increasing} \added{proliferating} number of artificial satellites is \deleted{correspondingly elevating} \added{increasing} the impact on observations with the 80-cm telescope and the MST at the Xinglong Observatory. This interference is more pronounced during \added{twilight and} summer. \deleted{ observations} The duration of satellite visibility is markedly greater in summer, leading to all night visibility, while in winter, visibility is confined to twilight. Particularly for wide-field telescopes like the MST. For the MST, the fraction of images containing satellite trails increased from 5\% in January 2023 to 12\% in December 2023, \deleted{this fraction reached} \added{with} a peak of 19\% \added{recorded} during the summer. \added{Furthermore, while
less than 20\% of twilight images were affected in January 2023, this fraction increased to approximately
35\% by December 2023.} For the 80-cm telescope, the fraction of images containing satellite trails  increased from an average of 0.34\% in 2019 to 0.7\% in 2023; \added{the impact is particularly severe during twilight}. Although the number of affected images \deleted{increases} \added{is rising} with increasing satellite deployments, \added{the} scientific operations \deleted{of the 80-cm telescope and the MST} have not yet been seriously compromised. However, \deleted{the number of satellites and space debris will only increase in the future} \added{as the population of satellites and space debris continues to grow, future interference will escalate}. We estimate that if the full constellation of 42,000 Starlink satellites is deployed, nearly all MST images \deleted{captured} \added{acquired} \deleted{in summer} \added{during twilight} \deleted{would} \added{will} be affected. \deleted{Once} \added{Should} the constellation scale reach 100,000 satellites, virtually all MST images \added{acquired in summer} will be affected. \deleted{regardless of the season} The fraction of sources affected by satellite trails remains \deleted{small} \added{modest} for both facilities. A clear peak in this fraction is observed during the summer.  

\added{Photometric analysis} \deleted{The results } demonstrates that sources \added{in closer} proximity to \deleted{the} satellite trails and those \deleted{that are intrinsically} \added{with} fainter \added{magnitudes} experience a greater impact. Specifically, \deleted{When the distance was 11 pixels from the trail} \added{at a distance of 11~pixels}, the median standardized residual \deleted{deviated} \added{deviates} by approximately $0.1\sigma$; \deleted{at a distance of 1 pixel, this deviation was} \added{this increases to} approximately $0.5\sigma$ \added{at 1~pixel}. \deleted{At a distance of 12 pixels, the} \added{The} median standardized residual approaches zero \added{at 12~pixels}, indicating that sources \deleted{farther than 12 pixels} \added{beyond this threshold} are virtually unaffected.
Regarding brightness, the median standardized residual progressively shifts toward negative values for sources fainter than \deleted{the 16th magnitude} \added{16~mag}, the deviation \deleted{increased} \added{grows} from approximately $0.15\sigma$ at \deleted{magnitude 16} \added{16~mag} to approximately $0.7\sigma$ at \deleted{magnitude 18} \added{18~mag}. \added{Furthermore, comparative analysis across different seeing conditions
shows that the deviation of median standardized residuals ($\sigma$) is significantly greater under
poor seeing than under good seeing conditions. Specifically, for the Poor Seeing group, the deviation of median standardized residuals remain greater than 0.5 $\sigma$ for all sources within 8 pixels of the trail; notably, the median standardized residuals in this group fail to return to the baseline even as the distance from the trail center increases to 12 pixels.}

Future work will \added{extend this analysis to} examine satellite interference with SiTian \citep{liu2021sitian} observations. As \deleted{more satellites are launched} \added{satellite populations continue to expand}, \deleted{observers} \added{it is imperative for the astronomical community to} develop advanced techniques to identify, model, subtract, and mask affected pixels. \deleted{in image processing} Concurrently, we call for \deleted{more decisive measures} \added{stringent regulatory frameworks} at the legislative and policy levels to preserve Dark and Quiet Skies, coupled with global scientific outreach to enhance \added{global} awareness of Earth and sky protection.

\begin{acknowledgements}
%This work was funded by the National Natural Science Foundation of China (NSFC) under No.11080922.
We acknowledge the support of the staff at the Xinglong Observatory. This work is supported by the Strategic Priority Research Program of the Chinese Academy of Sciences, Grant No. XDB0550100. Data resources are supported by China National Astronomical Data Center (NADC).

The SiTian project is a next-generation, large-scale time-domain survey designed to build an array of over 60 optical telescopes, primarily located at observatory sites in China. This array will enable single-exposure observations of the entire northern hemisphere night sky with a cadence of only 30-minute, capturing true color (gri) time-series data down to about 21 mag. This project is proposed and led by the National Astronomical Observatories, Chinese Academy of Sciences (NAOC). As the pathfinder for the SiTian project, the Mini-SiTian project utilizes an array of three 30 cm telescopes to simulate a single node of the full SiTian array. The Mini-SiTian has begun its survey since November 2022. The SiTian and Mini-SiTian have been supported from the Strategic Pioneer Program of the Astronomy Large-Scale Scientific Facility, Chinese Academy of Sciences and the Science and Education Integration Funding of University of Chinese Academy of Sciences.
\end{acknowledgements}

\bibliographystyle{raa}
\bibliography{bibtex}

@article{mcdowell2020low,
  title={The low earth orbit satellite population and impacts of the SpaceX Starlink constellation},
  author={McDowell, Jonathan C},
  journal={The Astrophysical Journal Letters},
  volume={892},
  number={2},
  pages={L36},
  year={2020},
  publisher={IOP Publishing}
}

@article{huang2025mini,
  title={The mini-SiTian array: a pathfinder for the SiTian Project},
  author={Huang, Yang and Liu, Jifeng and Wu, Hong and Shang, Zhaohui and Luo, Ali and Hu, Shaoming and Cui, Wenyuan and Mao, Yongna},
  journal={Research in Astronomy and Astrophysics},
  volume={25},
  number={4},
  pages={044001},
  year={2025},
  publisher={National Astromonical Observatories, CAS and IOP Publishing}
}

@article{zhang2025mini,
  title={The Mini-SiTian Array: Evaluation Camera System},
  author={Zhang, Yu and Du, Lin and Hu, Yi and Huang, Yang and Wu, Ying and He, Min and Gu, Hongrui and Mu, Haiyang and Chen, Xunhao and Wu, Hong},
  journal={Research in Astronomy and Astrophysics},
  volume={25},
  number={4},
  pages={044003},
  year={2025},
  publisher={National Astromonical Observatories, CAS and IOP Publishing}
}

@article{ref2,
  title={Impact of the SpaceX Starlink Satellites on the Zwicky Transient Facility Survey Observations},
  author={Mróz, Przemek and  Otarola, Angel  and  Prince, Thomas A.  and  Dekany, Richard  and  Duev, Dmitry A.  and  Graham, Matthew J.  and  Groom, Steven L.  and  Masci, Frank J.  and  Medford, Michael S. },
  journal={The Astrophysical Journal},
  volume={924},
  year={2022},
}

@article{cui2022impact,
  title={Impact simulation of Starlink satellites on astronomical observation using worldwide telescope},
  author={Cui, Z and Xu, Yunfei},
  journal={Astronomy and Computing},
  volume={41},
  pages={100652},
  year={2022},
  publisher={Elsevier}
}

@article{mallama2023assessment,
  title={Assessment of Brightness Mitigation Practices for Starlink Satellites},
  author={Mallama, Anthony and Hornig, Andreas and Cole, Richard E and Harrington, Scott and Respler, Jay and Lee, Ron and Worley, Aaron},
  journal={arXiv preprint arXiv:2309.14152},
  year={2023}
}

@article{mallama2023starlink,
  title={Starlink Generation 2 Mini satellites: photometric characterization},
  author={Mallama, Anthony and Cole, Richard E and Harrington, Scott and Hornig, Andreas and Respler, Jay and Worley, Aaron and Lee, Ron},
  journal={arXiv preprint arXiv:2306.06657},
  year={2023}
}

@article{xu2020ivoa,
  title={IVOA HiPS implementation in the framework of WorldWide Telescope},
  author={Xu, Yunfei and Cui, Chenzhou and Fan, Dongwei and Li, Shanshan and Li, Changhua and Han, Jun and Mi, Linying and He, Boliang and Yang, Hanxi and Tao, Yihan and others},
  journal={Astronomy and Computing},
  volume={31},
  pages={100380},
  year={2020},
  publisher={Elsevier}
}

@article{stoppa2024automated,
  title={Automated detection of satellite trails in ground-based observations using U-Net and Hough transform},
  author={Stoppa, F and Groot, PJ and Stuik, R and Vreeswijk, P and Bloemen, S and Pieterse, DLA and Woudt, PA},
  journal={Astronomy \& Astrophysics},
  volume={692},
  pages={A199},
  year={2024},
  publisher={EDP Sciences}
}

@article{krantz2021characterizing,
  title={Characterizing the all-sky brightness of satellite mega-constellations and the impact on astronomy research},
  author={Krantz, Harrison and Pearce, Eric C and Block, Adam},
  journal={arXiv preprint arXiv:2110.10578},
  year={2021}
}

@article{walker2020impact,
  title={Impact of satellite constellations on optical astronomy and recommendations toward mitigations},
  author={Walker, Constance and Hall, Jeffrey and Allen, Lori and Green, Richard and Seitzer, Patrick and Tyson, Tony and Bauer, Amanda and Krafton, Kelsie and Lowenthal, James and Parriott, Joel and others},
  journal={Bulletin of the American Astronomical Society},
  volume={52},
  number={2},
  year={2020}
}

@article{tyson2020mitigation,
  title={Mitigation of LEO satellite brightness and trail effects on the Rubin Observatory LSST},
  author={Tyson, J Anthony and Ivezi{\'c}, {\v{Z}}eljko and Bradshaw, Andrew and Rawls, Meredith L and Xin, Bo and Yoachim, Peter and Parejko, John and Greene, Jared and Sholl, Michael and Abbott, Timothy MC and others},
  journal={The Astronomical Journal},
  volume={160},
  number={5},
  pages={226},
  year={2020},
  publisher={IOP Publishing}
}

@article{zhi2024multicolour,
  title={Multicolour photometry of LEO mega-constellations Starlink and OneWeb},
  author={Zhi, Hui and Jiang, Xiaojun and Wang, Jianfeng},
  journal={Monthly Notices of the Royal Astronomical Society},
  volume={530},
  number={4},
  pages={5006--5015},
  year={2024},
  publisher={Oxford University Press}
}

@article{lawler2021visibility,
  title={Visibility predictions for near-future satellite megaconstellations: latitudes near 50 will experience the worst light pollution},
  author={Lawler, Samantha M and Boley, Aaron C and Rein, Hanno},
  journal={The Astronomical Journal},
  volume={163},
  number={1},
  pages={21},
  year={2021},
  publisher={IOP Publishing}
}

@article{gallozzi2020concerns,
  title={Concerns about ground based astronomical observations: quantifying satellites' constellations damages},
  author={Gallozzi, Stefano and Paris, Diego and Scardia, Marco and Dubois, David},
  journal={arXiv preprint arXiv:2003.05472},
  year={2020}
}

@article{bassa2022analytical,
  title={Analytical simulations of the effect of satellite constellations on optical and near-infrared observations},
  author={Bassa, CG and Hainaut, OR and Galad{\'\i}-Enr{\'\i}quez, David},
  journal={Astronomy \& Astrophysics},
  volume={657},
  pages={A75},
  year={2022},
  publisher={EDP Sciences}
}

@article{he2025mini,
  title={The Mini-SiTian Array: first-two-year operation},
  author={He, Min and Wu, Hong and Ge, Liang and Tian, Jianfeng and Wang, Zheng and Mu, Haiyang and Zhang, Yu and Huang, Yang and Zheng, Jie and Fan, Zhou and others},
  journal={Research in Astronomy and Astrophysics},
  volume={25},
  number={4},
  pages={044005},
  year={2025},
  publisher={IOP Publishing}
}

@article{fan2016,  
title={The Xinglong 2.16-m Telescope: Current Instruments and Scientific Projects},  
author={ Fan, Zhou  and  Wang, Huijuan  and  Jiang, Xiaojun  and  Wu, Hong  and  Li, Hongbin  and  Huang, Yang  and  Xu, Dawei  and  Hu, Zhongwen  and  Zhu, Yinan  and  Wang, Jianfeng  and },  
journal={Publications of the Astronomical Society of the Pacific},  
number={969},  
pages={128},  
year={2016},
}

@article{kruk2023HST,
  title={The impact of satellite trails on Hubble Space Telescope observations},
  author={Kruk, Sandor and Garc{\'\i}a-Mart{\'\i}n, Pablo and Popescu, Marcel and Aussel, Ben and Dillmann, Steven and Perks, Megan E and Lund, Tamina and Mer{\'\i}n, Bruno and Thomson, Ross and Karadag, Samet and others},
  journal={Nature Astronomy},
  volume={7},
  number={3},
  pages={262--268},
  year={2023},
  publisher={Nature Publishing Group UK London}
}

@article{muirhead2025modeling,
  title={Modeling the Optical Impact of the Second-generation Starlink Satellite Constellation},
  author={Muirhead, Ian J and Crisp, Nicholas H and McGrath, Ciara N and Roberts, Peter CE},
  journal={The Astronomical Journal},
  volume={170},
  number={4},
  pages={215},
  year={2025},
  publisher={IOP Publishing}
}

@article{tregloan2020first,
  title={First observations and magnitude measurement of Starlink’s Darksat},
  author={Tregloan-Reed, J and Otarola, A and Ortiz, E and Molina, V and Anais, J and Gonz{\'a}lez, R and Colque, JP and Unda-Sanzana, E},
  journal={Astronomy \& Astrophysics},
  volume={637},
  pages={L1},
  year={2020},
  publisher={EDP Sciences}
}

@article{halferty2022photometric,
  title={Photometric characterization and trajectory accuracy of Starlink satellites: implications for ground-based astronomical surveys},
  author={Halferty, Grace and Reddy, Vishnu and Campbell, Tanner and Battle, Adam and Furfaro, Roberto},
  journal={Monthly Notices of the Royal Astronomical Society},
  volume={516},
  number={1},
  pages={1502--1508},
  year={2022},
  publisher={Oxford University Press}
}

@article{liu2021sitian,
  title={The sitian project},
  author={Liu, Jifeng and Soria, Roberto and Wu, Xue-Feng and Wu, Hong and Shang, Zhaohui},
  journal={Anais da Academia Brasileira de Ci{\^e}ncias},
  volume={93},
  number={suppl 1},
  pages={e20200628},
  year={2021},
  publisher={SciELO Brasil}
}

@article{barentine2023aggregate,
  title={Aggregate effects of proliferating low-Earth-orbit objects and implications for astronomical data lost in the noise},
  author={Barentine, John C and Venkatesan, Aparna and Heim, Jessica and Lowenthal, James and Kocifaj, Miroslav and Bar{\'a}, Salvador},
  journal={Nature Astronomy},
  volume={7},
  number={3},
  pages={252--258},
  year={2023},
  publisher={Nature Publishing Group UK London}
}

@article{horiuchi2020simultaneous,
  title={Simultaneous multicolor observations of Starlink’s Darksat by the Murikabushi Telescope with MITSuME},
  author={Horiuchi, Takashi and Hanayama, Hidekazu and Ohishi, Masatoshi},
  journal={The Astrophysical Journal},
  volume={905},
  number={1},
  pages={3},
  year={2020},
  publisher={IOP Publishing}
}

@article{rosenfield2018aas,  
title={AAS WorldWide telescope: a seamless, cross-platform data visualization engine for astronomy research, education, and democratizing data},  
author={Rosenfield, Philip and Fay, Jonathan and Gilchrist, Ronald K and Cui, Chenzhou and Weigel, A David and Robitaille, Thomas and Otor, Oderah Justin and Goodman, Alyssa},  
journal={The Astrophysical Journal Supplement Series},  
volume={236},  
number={1},  
pages={22},  
year={2018},  
publisher={IOP Publishing}
}
\label{lastpage}

\end{document}